\documentclass[12pt]{article}
\usepackage{graphics}
\usepackage{bm}
\usepackage{graphicx}
\usepackage{amsmath}
\usepackage{amssymb}
\usepackage{amstext}
\usepackage{amscd}
\usepackage{amsfonts}
\usepackage{appendix}
\usepackage{color}
\usepackage{wasysym}
\usepackage{latexsym}
\DeclareMathAlphabet{\EuFrak}{U}{euf}{m}{n}
\DeclareMathAlphabet{\EuScript}{U}{eus}{m}{n}

\newcommand{\nd}{\noindent}

\newcommand{\be}{\begin{equation}}
\newcommand{\ee}{\end{equation}}
\newcommand{\ben}{\begin{eqnarray}}
\newcommand{\een}{\end{eqnarray}}

\title{{\bf A tale of two probability distributions} \\ {\small {\it Reply to Comment on ''Troublesome aspects of the Renyi-MaxEnt treatment” by Thomas Oikonomou and G. Baris Bagci} }}
\author{{A. Plastino$^{1,3,4}$, M.C.Rocca$^{1,2,3}$}, F. Pennini$^{5,6}$ \\
\small{$^1$ Departamento de F\'{\i}sica,
Universidad Nacional de La Plata,}\\
\small{$^2$ Departamento de Matem\'{a}tica,
Universidad Nacional de La Plata,}\\
\small{$^3$ Consejo Nacional de Investigaciones Cient\'{\i}ficas
y Tecnol\'{o}gicas}\\
\small{(IFLP-CCT-CONICET)-C. C. 727, 1900 La Plata -
Argentina}\\\small{$^4$  SThAR - EPFL, Lausanne, Switzerland}\\
\small{$^{5}$Universidad Cat\'olica del Norte, Av.~Angamos~0610, 
Antofagasta, Chile.}\\
\small{$^{6}$Facultad de Ciencias Exactas y Naturales,}\\
\small{Universidad Nacional de La Pampa, Peru 151, 6300 Santa Rosa,}\\
\small{La Pampa, Argentina}}

\date{\today}

\begin{document}

\maketitle

\begin{abstract}
\noindent 
This Reply is intended as a refutation of the preceding Comment [Oikonomou and Bagci, Phys. Rev. E
96, 056101 (2017)] on our paper [Plastino et al., Phys. Rev. E 94, 012145 (2016).]. We show that the Tsallis
probability distribution of our paper does not coincide with the Tsallis distribution studied by Oikonomou and
Bagci. Consequently, their findings do not apply to our paper.\\
DOI: 10.1103/PhysRevE.96.056102

\end{abstract}

\section{Introduction}

\nd The authors, OB for short, are here criticizing our PRE paper 
cited in their reference \cite{[1]}. Let us start with three statements regarding the essence of the discussion in [1].
\vskip 3mm
\nd A1:  It is shown in [1] that the MaxEnt variational approach used in conjunction with Renyi's entropy leads to inconsistencies. 
\vskip 3mm
\nd A2: These inconsistencies are due to a hidden relation between the concomitant Legendre multipliers 
($\lambda_1$ and $\lambda_2$) discovered while dissecting the variational process that leads to the appropriate   MaxEnt  probability distribution.
\vskip 3mm
\nd A3: Renyi's entropy is not trace form, while Tsallis' one is of such nature. Thus, we can expect differences to arise in a MaxEnt treatment.
\vskip 3mm
\nd
We pass now to the essence of the Comment:
\vskip 3mm
\nd B1: OB do not question the first two above points. Their claim is that they apply also to Tsallis' entropy. More precisely, they purport to discover that a hidden relation emerges in the Tsallis' MaxEnt treatment as well.

\section{The two-distributions' problem}

\nd OB work with the pseudo-Tsallis probability distribution (PD)

\begin{equation}
\label{1eq4}
P_{OB}=Z^{-1}
\left[1+(1-q)
\beta Z^{q-1}(U-<U>)\right]^\frac {1} {q-1}
\end{equation}

\begin{equation}
\label{1eq3}
Z=\left(\int P_{OB}^qd\mu\right)^{\frac {1} {1-q}},
\end{equation}
which does not coincide with Tsallis' pioneer PD of 1988 \cite{Tbook},
which reads

\begin{equation}
\label{1eq2.22} P_T=\frac {[1+\beta(1-q)U]^{\frac {1} {q-1}}} {Z_T},
\end{equation}
\begin{equation}
\label{1eq2.23} Z_T=\int\limits_M [1+\beta(1-q)U]^{\frac {1}
{q-1}}\;d\mu.
\end{equation}
Thus, we can rephrase B1 as stating: \newline \nd \nd B1: OB purport to discover that a hidden relation emerges in the Tsallis' MaxEnt treatment for their PD $P_{OB}$, which is not Tsallis' PD. We coul finish our Reply right here. However, let us delve deeper into the issue in order to gain some insight into why OB get their peculiar PD distribution (\ref{1eq4}). Such is the subject of next Section.

\section{Boltzmann-Gibbs \`a  la Oikonomou-Bagci}

\subsection{Normal procedure}

\nd In order to better illustrate the OB procedure, we apply it here to the 
Boltzmann-Gibbs (BG)  exponential distribution.  
One maximizes in such an instance
\begin{equation}
\label{ep3.1}
F_{S_B}(P)=-\int P\ln P d\mu+\lambda_1\left(\int PU d\mu-<U>\right)+
\lambda_2\left(\int Pd\mu-1\right).
\end{equation}
The first variation becomes
\begin{equation}
\label{ep3.2}
F_{S_B}(P+h)-F_{S_B}(P)=-\int\left(\ln P-\lambda_1U-
\lambda_2+1\right)hd\mu+O(h^2).
\end{equation}
Accordingly,  
\begin{equation}
\label{ep3.3}
\ln P-\lambda_1U-\lambda_2+1=0.
\end{equation}
Here, as most people do, {\it but not OB}, one immediately deduces $P$ and  is immediately led to 
\begin{equation}
\label{ep3.4}
P=e^{\lambda_1U}e^{\lambda_2-1}.
\end{equation}
$P-$normalization entails
\begin{equation}
\label{ep3.5}
e^{\lambda_2-1}\int e^{\lambda_1U}d\mu=1,
\end{equation}
and then
\begin{equation}
\label{ep3.6}
e^{\lambda_2-1}=\frac {1} {\int e^{\lambda_1U}d\mu}.
\end{equation}
In other words, 
\begin{equation}
\label{ep3.7}
P=\frac {e^{\lambda_1U}} {\int e^{\lambda_1U}d\mu}.
\end{equation}
Further, one finds
\begin{equation}
\label{ep3.8}
<U>=\frac {\int Ue^{\lambda_1U}d\mu} {\int e^{\lambda_1U}d\mu}.
\end{equation}
Well known physical arguments, as shown first by Gibbs himself \cite{gibbs}, allow one to identify 
$\lambda_1$ 
\be 
\label{ep3.9} 
\lambda_1=-\beta= -\frac{1}{kT}.\ee

\subsection{OB procedure}
Starting with Eq. (\ref{ep3.3}), OB follow a different trajectory so as to ascertain which is proper $P$. 
They first multiply  (\ref{ep3.3}) by $P$ and integrate, finding

\begin{equation}
\label{ep3.10}
\int P\ln P d\mu-\lambda_1<U>-\lambda_2+1=0,
\end{equation}
so that

\begin{equation}
\label{ep3.11}
\lambda_2=\int P\ln P d\mu+\beta<U>+1,
\end{equation}
which OB  would call a {\it hidden relation between $\lambda_2$ and $\beta$}. Replacing (\ref{ep3.11}) into (\ref{ep3.3}) OB obtain
\begin{equation}
\label{ep3.12}
\ln P+\beta(U-<U>)-\int P\ln P d\mu=0,
\end{equation}
or
\begin{equation}
\label{ep3.13}
P=e^{-\beta(U-<U>)+\int P\ln P d\mu}.
\end{equation}
 Integrating once again OB are led to
\begin{equation}
\label{ep3.14}
e^{\int P\ln P d\mu}\int e^{-\beta(U-<U>)} d\mu=1.
\end{equation}
This is a critical stage. OB choose to  write $Z$ not as 
\begin{equation}
\label{ep3.15}
{\cal Z}^{-1}=e^{\int P\ln P d\mu+\beta<U>}, 
\end{equation}
but as 

\begin{equation}
\label{ep3.19}
{\cal Z_{OB}}^{-1}=e^{\int P_{OB}\ln P_{OB} d\mu}, 
\end{equation}
leading to 
\begin{equation}
\label{ep3.20}
P_{OB}=\frac {e^{\beta(U-<U>)}} {{\cal Z_{OB}}},
\end{equation}
and then it follows that 
\begin{equation}
\label{ep3.21}
{\cal S}=\ln {\cal Z_{OB}},
\end{equation}
which is obviously an incorrect result. This happens because of the OB-choice (\ref{ep3.19}). 
 With the selection (\ref{ep3.15}) they would have reached instead 
\begin{equation}
\label{ep3.16}
P=\frac {e^{-\beta U}} {{\cal Z}},
\end{equation}
so that
\begin{equation}
\label{ep3.17}
e^{-\int P\ln P d\mu} ={\cal Z}e^{\beta<U>},
\end{equation}
and

\begin{equation}
\label{ep3.18}
{\cal S}=\ln {\cal Z}+\beta<U>,
\end{equation}
the correct result. We have clearly identified, with reference tp the BG distribution, the origin of OB's troubles.

\section{ Tsallis' PD}

\subsection{Normal procedure}

The first variation's equation is [1]
\begin{equation}
\label{ep4.1}
\frac {q} {1-q}P^{q-1}+\lambda_1U+\lambda_2=0
\end{equation}
\begin{equation}
\label{ep4.2}
\lambda_1=-\beta q{\cal Z}_q^{1-q}\;\;\;
\lambda_2=\frac {q} {q-1}{\cal Z}_q^{1-q},
\end{equation}
leading to
\begin{equation}
\label{ep4.3}
P=\frac {[1+(1-q)\beta U]^{\frac {1} {q-q}}} {{\cal Z}_q},
\end{equation}
and, for ${\cal S}_q$,
\begin{equation}
\label{ep4.4}
{\cal S}_q=\ln_q{\cal Z}_q+{\cal Z}_q^{1-q}\beta<U>,
\end{equation}
the correct result.

\subsection{ Tsallis' PD  \`a la Oikonomou-Bagci}

OB multiply (\ref{ep4.1}) by $P$ and  integrate:

\begin{equation}
\label{ep4.5}
\frac {q} {1-q}\int P^qd\mu+\lambda_1<U>+\lambda_2=0,
\end{equation}
which they call a {\it hidden relation} between two Lagrange multipliers entering the MaxEnt treatment. 
This is equation (1) in the Comment, the hard core of their present contribution. They now choose
\begin{equation}
\label{ep4.6}
\lambda_1=-\frac {\beta q\int P^qd\mu} {1+(1-q)\beta<U>},
\end{equation}
and  obtain 
\begin{equation}
\label{ep4.7}
\lambda_2=\frac {q} {q-1}\int P^qd\mu+\frac {\beta q\int P^qd\mu} {1+(1-q)\beta<U>},
\end{equation}
so that 
\begin{equation}
\label{ep4.8}
P^{q-1}=\frac {\beta q\int P^qd\mu} {1+(1-q)\beta<U>}[1+(1-q)\beta U],
\end{equation}
or
\begin{equation}
\label{ep4.9}
P=\left(\frac {\beta q\int P^qd\mu} {1+(1-q)\beta<U>}\right)^{\frac {1} {q-1}}.
[1+(1-q)\beta U]^{\frac {1} {q-1}}.
\end{equation}
OB are here at a critical stance. Had they selected
\begin{equation}
\label{ep4.10}
{\cal Z}_q=\left(\frac {\beta q\int P^qd\mu} {1+(1-q)\beta<U>}\right)^{\frac {1} {1-q}},
\end{equation}
 they would have found for $P$ the expression (\ref{ep4.3}), the right Tsallis'  result, 
obtained the hidden relation (\ref{ep4.5}) notwithstanding. We see that, contrary to OB's claim, the hidden relation impedes nothing. However, 
at this  crucial  stage  OB chose to write 
\begin{equation}
\label{ep4.11}
\lambda_1=-\beta q.
\end{equation}
leading to 
\begin{equation}
\label{ep4.12}
P=\left(\int P^qd\mu\right)^{\frac {1} {q-1}}
\left[1+\frac {1+(1-q)\beta(U-<U>)} {\int P^qd\mu}\right]^{\frac {1} {q-1}},
\end{equation}
which is Eq. (1) above.  
According to the OB-choice (\ref{ep4.11}) above we have now
\begin{equation}
\label{ep4.13}
{\cal Z}_q=\left(\int P^qd\mu\right)^{\frac {1} {1-q}},
\end{equation}
and 
\begin{equation}
\label{ep4.14}
{\cal S}_q=\ln_q{\cal Z}_q,
\end{equation}
an incorrect result, arising because $\lambda_1$ was incorrectly chosen.

\section{MaxEnt reciprocity relations}

\nd A word of caution. We have above used words like ''choosing'' or ''selection''. This is speaking in a rather loose fashion. In fact, MaxEnt prescribes a definite recipe in order to find the Lagrange multipliers  $\lambda$.
 MaxEnt asserts that, if the a priori known information concerns $N$ 
expectation values \newline $<A_k>$, and then $N+1$ (accounting for normalization) Lagrange multipliers $\lambda_k$, then the entropy $S$ acquires the form (MaxEnt version of $S$)

\be S= \lambda_0 +  \sum_{k=1}^N\, \lambda_k <A_k>.   \ee
The $\lambda_k$'s are obtained via so-called reciprocity relations  (see, for example, \cite{flego})
\be \lambda_k=  \frac{\partial S}{\partial <A_k>}. \label{uuu} \ee
 In practice, however, instead of solving equations 
(\ref{uuu}) one often makes educated guesses for the  $\lambda_k$'s, 
as reported above in this paper. \vskip 3mm

How to make such an educated guess in the Tsallis' instance? A main criterion is to choose  
Tsallis' $\lambda_1$ in such a manner that, in the limit $q\rightarrow 1$,  it should coincide with the 
 Boltzmann-Gibbs' $\lambda_1$. In that case, from such correct $\lambda_1$ one immediately derives a Tsallis' $\lambda_2$, 
   that yields then the usual Tsallis' distribution.  This $\lambda_1$-criterion is satisfied by both OB's guess (31) and Tsallis' one (27). One can appeal then to Ockham's razor to select (27). 	A word of caution seems appropriate. Guessing is an art, not science. If inspiration fails in guessing the Lagrange parameter, one can always appeal to Eq. (41), that never fails.

\section{Conclusion}
The Comment to which we have replied here  serves the useful purpose of highlighting issues related to  Tsallis' statistics, but does not invalidate our paper [1].  OB are not using Tsallis' PD distribution but one concocted by them. The Comment's main result is its  Eq. (1), which we showed does not impede one to arrive at the correct Tsallis' PD. The Comment's error lies in a not judicious choice of Lagrange multipliers. This invalidates its conclusions, fruits of the poisoned tree.

\end{document}